\documentclass[useAMS,usenatbib]{mn2e}
\usepackage{natbib}

\newcommand*{\Msun}{\ensuremath{\mathrm{M_\odot}}}%
\newcommand*{\Lsun}{\ensuremath{\mathrm{L_\odot}}}%
\newcommand*{\Mpc}{\ensuremath{\mathrm{Mpc}}}%

\usepackage{graphicx}  
\usepackage{epstopdf}
\title[Star Formation in the GOODS NICMOS Survey (GNS)]{Star Formation in a Stellar Mass Selected Sample of Galaxies to $z=3$ from the GOODS NICMOS Survey}
\author[A. E. Bauer et al.]
{A.~E.~Bauer,$^{1,2}$\thanks{E-mail:
abauer@aao.gov.au}
C.~J.~Conselice,$^{2}$
P.~G.~P\'erez-Gonz\'alez,$^{3,4}$ 
R.~Grutzbauch,$^{2}$
\newauthor  A.~F.~L.~Bluck,$^{5}$
F.~Buitrago,$^{2}$
A.~Mortlock,$^{2}$\\
$^{1}$Australian Astronomical Observatory, PO Box 296, Epping, NSW 1710, Australia\\
$^{2}$University of Nottingham, School of Physics and Astronomy, Nottingham NG1 3AL, UK\\
$^{3}$Departamento de Astrof\'isica, Facultad de CC. F\'isicas, Universidad Complutense de Madrid, E-28040 Madrid, Spain\\
$^{4}$Associate Astronomer at Steward Observatory, The University of Arizona, USA\\
$^{5}$Gemini Observatory, Hilo, Hawaii, 96720, USA\\
}

\begin{document}

\date{accepted to MNRAS: 13 June 2011}

\pagerange{\pageref{firstpage}--\pageref{lastpage}} \pubyear{2011}

\maketitle

\label{firstpage}

\begin{abstract}
We present a study of the star-forming properties of a stellar mass-selected sample of galaxies in the GOODS NICMOS Survey (GNS), based on deep Hubble Space Telescope imaging of the GOODS North and South fields.   Using a stellar mass selected sample, combined with HST/ACS and {\rm Spitzer} data to measure both UV and infrared derived star formation rates, we investigate the star forming properties of a complete sample of $\sim$1300 galaxies down to log~$M_{*}=9.5$ at redshifts $1.5<z<3$.  Eight percent of the sample is made up of massive galaxies with $M_{*} \ge10^{11}\Msun$.  We derive optical colours, dust extinctions, and ultraviolet and infrared star formation rates (SFR) to determine how the star formation rate changes as a function of both stellar mass and time.   Our results show that SFR increases at higher stellar mass such that massive galaxies nearly double their stellar mass from star formation alone over the redshift range studied, but the average value of SFR for a given stellar mass remains constant over this $\sim2$ Gyr period.   Furthermore, we find no strong evolution in the SFR for our sample as a function of mass over our redshift range of interest, in particular we do not find a decline in the SFR among massive galaxies, as is seen at $z < 1$.  The most massive galaxies in our sample (log $M_{*} \ge11$) have high average star formation rates with values, SFR$_{UV,corr}$ = $103\pm75~\Msun~$yr$^{-1}$, yet exhibit red rest-frame $(U-B)$ colours at all redshifts.  We conclude that the majority of these red high-redshift massive galaxies are red due to dust extinction.  We find that A$_{2800}$ increases with stellar mass, and show that between 45\% and 85\% of massive galaxies harbour dusty star formation.  These results show that even just a few Gyr after the first galaxies appear, there are strong relations between the global physical properties of galaxies, driven by stellar mass or another underlying feature of galaxies strongly related to the stellar mass.
     
\end{abstract}


\section{Introduction}

One of the least understood aspects of galaxy formation and evolution is the triggering and regulation of star formation in galaxies, and how this varies across galaxy mass and time.  Galaxies are formed in star formation events, and understanding how and when star formation occurs is critical for understanding the evolution of galaxies.  One intriguing aspect of the star formation in galaxies is that the star formation history appears to vary strongly as a function of galaxy mass, at least at $z < 1.5$.  Even within the local universe we observe that the most massive galaxies are quiescent ellipticals, while lower mass galaxies are undergoing significant star formation in either irregular or spiral galaxies \citep[e.g.][]{conselice2006}.    One key way to address galaxy formation and evolution directly is to understand how the nearby galaxy population was put into place and evolved from higher redshift galaxies which we can now observe in nearly complete mass-selected samples up to $z =3$ \citep[e.g.][]{Daddi07,Conselice2011}.
 
Observational results at redshifts $z < 1.4$ reveal that the most massive galaxies ($M_{*} \ge10^{11}\Msun$) appear to be dominated by old stellar populations with negligible amounts of star formation, and are largely assembled by $z\sim1$ \citep{McCarthy04,Daddi05,Saracco05,Labbe06,Conselice2007}.  These galaxies therefore must have formed very quickly inside of matter over-densities, within a few billion years after the Big Bang, then stopped forming stars, presumably as cold gas supplies diminished.  This is in broad agreement with analyses of stellar populations seen in nearby massive ellipticals \citep[e.g.][]{Zhu2010}, and is consistent with the broad trends seen in colours \citep[e.g.][]{Faber2007,Kriek2008}.

Following up on this, several studies have shown that star formation rates (SFRs) correlate with stellar mass \citep{Daddi07,noeske07,Pannella2009} up to $z=2$ and that the most massive galaxies appear to shut off their star formation at earlier times than  low-mass galaxies \citep[e.g.][]{CSHC96,BE00,Bauer05,Feulner05,PerezGonzalez05,Juneau05,Bundy06,Papovich2006,Zheng07,Schiminovich07,Damen2009,Bauer2010}.   This effect of star formation ending in higher mass galaxies before lower mass systems is known as ``downsizing.''  The origin of downsizing is still not understood and must be the result of the removal of cold gas from a galaxy or the heating of this gas by some feedback process.   Possibilities for the origin of this feedback include:  active galactic nuclei (AGN), supernova, energy from a merger, or a combination of these.    Standard galaxy formation models have difficulty reproducing downsizing as well as the properties of redder galaxies observed at high redshifts \citep[e.g.][]{Cole2000}, and properties of modern massive red galaxies that harbour old stellar populations.  

Furthermore, we know very little about how star formation and galaxy mass assembly rates vary as a function of galaxy mass and time at high redshift. While there is evidence that galaxy mergers vary with stellar mass at high redshifts, such that the most massive galaxies are undergoing the most merging at $z > 1$ \citep[e.g.][]{CBDP03,Conselice2008,Bluck2009}, it is largely unknown how other modes of formation, such as cold gas accretion or in-situ star formation, are driving the formation of galaxies at early times.  A deep, and at least moderately large area, survey complete in stellar mass is needed to address this point.

There exists some evidence for how massive galaxies at high redshifts form stars, and the nature of their stellar population content, based on their optical and near infrared colours and limited spectroscopy at $z\sim2$.  Massive galaxies in the distant universe have been observed to be often very red, creating what is possibly the origin of the ``red sequence'' found in nearby massive galaxies \citep[e.g.][]{Conselice2007,Kriek2008}.  Usually it is assumed that these red galaxies seen in the distant universe consist of old stellar populations.  However, it is possible that these massive galaxies are red due to dust, or a combination of older stellar populations and dusty star formation.  It is therefore possible that the dust content of massive galaxies in the universe has been underestimated, and a full consensus of the dust properties of high redshift massive galaxies is not yet known.  

There is evidence that dusty star formation is common in producing red colours in lower redshift galaxies at $z < 1$.   For example, by using a $K$-selected sample of galaxies, \citet{K20-02} found that among $K_s<19.2$ Extremely Red Objects (EROs) at $z\sim1$, 50\% are old passively evolving galaxies while the other half are dusty star-forming galaxies.  In a more recent study of EROs at $z\sim1$, \citet{Kong2008} confirm these results, showing a passive fraction of $\sim$48\% at $K < 19.2$, in an otherwise active star-forming population.

There is also evidence that more distant galaxies are in fact redder due to dusty star formation, rather than from old stellar populations.  For example,  \citet{Pannella2009} examine a sample of  star-forming BzK galaxies from the COSMOS survey at $z > 1.5$ and derive dust attenuation factors from both UV and radio wavelengths.  They find a strong agreement between the two methods of measuring dust content, and also show that for their sample of galaxies with median redshift $z\sim1.7$, dust extinction increases with stellar mass.  However, what is missing from these studies is a stellar-mass-selected sample of galaxies to systematically investigate these trends, and to determine how the star formation rate is changing with stellar mass and redshift.  Furthermore, we are also interested in further investigating the role of dusty star formation in the observed trend that massive galaxies are red at high redshift, while lower mass systems tend to be bluer \citep{Grutzbauch2011}.

One  way to address these issues is to directly probe the star formation rate evolution of high redshift galaxies as a function of stellar mass and time.  This allows us to determine directly the assembly rate for these systems at early times, and how this mode of assembly varies as a function of the already-established relationship with stellar mass.  This has been done to some extent \citep{Daddi07,noeske07}, but has been largely limited to examining the most massive galaxies, and furthermore only a subset of these which are star forming.

In this paper, we study the properties of a mass-selected, complete sample of galaxies from the GOODS NICMOS Survey (GNS) \citep{Conselice2011}, focusing the analysis on star-forming properties of galaxies with log~$M_{*}\ge 9.5$ in the redshift range of $1.5<z<3.0$.  In $\S~\ref{sec:data}$ we describe the observations and sample selection.    We describe our methods for determining physical properties of the galaxies in $\S~\ref{sec:sfrall}$, including a discussion of the reliability of SFR indicators in this redshift range.  Section $\ref{sec:results}$ discusses the trends of star-forming properties with colour, stellar mass, and redshift.  We summarise our conclusions in $\S~\ref{sec:conclusions}$.  Throughout this work we adopt an $\Omega_M = 0.3$, $\Omega_{\Lambda} = 0.7$, $H_0 = 70\ \mathrm{km\ s^{-1} \Mpc^{-1}}$ cosmology, use a Salpeter initial mass function, and provide all magnitudes in the AB system unless otherwise noted.

\section[]{Data and Sample}\label{sec:data}

In this section we introduce the data we use from the GOODS-NICMOS Survey (GNS) and describe the derivation of photometric redshifts and stellar masses.  We then discuss the completeness of the survey at the redshifts of the study and finally introduce the supplementary infrared data. 

\subsection{The GOODS-NICMOS Survey}\label{sec:gns}

Our galaxy sample comes from the GOODS-NICMOS Survey \citep[GNS;][]{Conselice2011}.  The GNS is a large Hubble Space Telescope program conducted with the NICMOS NIC3 instrument, to obtain deep $H$-band and $J$-band imaging within the GOODS North and South fields \citep[GOODS: Great Observatories Origins Deep Survey;][]{Dickinson2003}.  Using 180 HST orbits, a total of 60 pointings were observed, centered on the most massive galaxies ($M_{*} \ge10^{11}\Msun$) at $1.7<z<2.9$.  The massive galaxies on which the pointings are centred were selected based on previous colour-selected samples within the GOODS fields that were designed to find distant galaxies, including $BzK$-selected galaxies, Extremely Red Objects (EROs), and Distant Red Galaxies (DRGs).  While the primary field selection was done in terms of the massive galaxy selection through these three primary colour criteria, to optimise the field placement, catalogues of Lyman-break BM/BX objects as well as high redshift drop-outs and sub-mm galaxies, were also used.  For a full description of the sample selection see \citet{Conselice2011}.

The main observations for the GNS survey were carried out using the NICMOS-3 camera on HST to obtain 3 orbits per pointing for each image.  With this exposure time, we achieved a F160W ($H$-band) depth of 26.8 mag (AB) at 5$\sigma$ using a 0.7 arcsec-diameter aperture.    The field of view of each image is 51.2 arcsec on a side with a pixel scale of 0.1 arcsec pixel$^{-1}$.  The FWHM for our images is roughly 0.3 arcsec.  The data reduction is described in detail by \citet{MBI2007} who detect sources in the $H$-band using {\tt SExtractor} \citep{BA1996}.  The full GNS catalog is derived to include every non-stellar object identified in the $H$-band imaging over the total area of the survey: 45~arcmin$^2$.  Overall, we find a total of 8298 objects in the GNS area.  

We supplement the near-infrared observations by matching sources to the GOODS/ACS optical catalog \citep{Giavalisco04}.   Photometry in $B, V, i$ and $z$ is available for sources down to a limiting magnitude of $B \sim 28.2$.  After accounting for the known 0.3 arcsec offset in declination in the ACS v2 catalog, we perform source matching between the NICMOS/$H$-band and ACS/$z_{850}$-band catalogs.  The mean separation between optical and $H$-band coordinates their ACS counterpart is $\sim 0.07 \pm 0.15 ^{\prime\prime}$ for both the North and South fields.  For the original matching between samples, a radius of up to 2$^{\prime\prime}$ was allowed, but in the resulting sample, only 18 of the matches have a separation greater than 0.5$^{\prime\prime}$.  We do not find a strong correlation between catastrophic failures in photometric redshift and having a matching radius separation greater than 1 arcsec, or 1.5 arcsec separation so we have left these objects in the sample.


\label{sec:properties}

\subsection{Redshifts}\label{sec:redshifts}

Where possible, we use spectroscopic redshifts published in the literature for GNS galaxies, otherwise we use our own measured photometric redshifts.  Spectroscopic redshifts of sources in the GOODS-N field were compiled by \citet{Barger2008}, whereas the GOODS-S field spectroscopic redshifts are taken from the FIREWORKS compilation \citep{Wuyts2008}. We matched these catalogs to the full GNS photometric catalog with a matching radius of $2 ^{\prime\prime}$, finding very few multiple matches, and a mean separation between photometric and spectroscopic sources of $0.11 \pm 0.06^{\prime\prime}$ in the GOODS-N field and $0.13 \pm 0.05^{\prime\prime}$ in the GOODS-S field.  In the full GNS sample, there are 537 spectroscopic redshifts for sources in GOODS-N and 369 in GOODS-S.    

For the remaining GNS galaxies, photometric redshifts were obtained by fitting template spectra to the $BVizH$ photometric data points, following the standard $\chi ^2$ minimization procedure using HYPERZ \citep{Bolzonella2000} as described fully in \citet{Grutzbauch2011} and \citet{Conselice2011}.  The synthetic spectra used by HYPERZ are constructed with the \citet{BC03} evolutionary code representing roughly the different morphological types of galaxies found in the local universe. We use five template spectra corresponding to the spectral types of E, Sa, Sc and Im as well as a single burst scenario and a reddening law taken from \citet{Calzetti00}.  We compute the most likely redshift solution and corresponding probability in the parameter space of age, metallicity and reddening.

With the filters we use to determine photometric redshifts the reddest rest-frame wavelength we probe is $B$-band at $z\sim3$.  We cannot use longer wavelength observations to determine photometric redshifts because there are no near-IR observations redder than $H$-band that have the same fidelity and depth as the $BVizH$ bands we use in this study.  The ground based $K$-band observations available for this field are not nearly as deep as our NICMOS $H$-band observations.  While we have IRAC data for our sources, we do not use these data to determine photometric redshifts or stellar masses due to issues with the PSF and contamination from neighbouring galaxies.  Furthermore, the rest-frame B-band gives us a good anchor for measuring stellar masses, as is shown by \citet[e.g.][]{BD01} at lower redshifts, and \citet{Bundy06} for higher redshifts.

As with any wide-field study of galaxies at $z>1$, the accuracy of photometric redshifts presents a challenge when interpreting observational results and cannot be neglected.  We define the reliability of photometric redshift measures as $\Delta z/(1+z) \equiv (z_{spec}- z_{phot})/(1+z_{spec})$.  We compare the average error ($\langle \Delta z/(1+z) \rangle$) and rms scatter ($\sigma$) as well as the fraction of catastrophic outliers, i.e., galaxies with $|\Delta z/(1+z)| > 0.3$.
Using HYPERZ, we find that sources with a high probability ($P > 95\%$) have $\langle \Delta z/(1+z) \rangle = 0.03$, with a scatter of $\sigma = 0.045$ (356 out of 906 galaxies with $P > 95\%$). Considering all sources regardless of their probability we find similar values: $\langle \Delta z/(1+z) \rangle = 0.01$ and $\sigma = 0.06$. We therefore consider all photometric redshifts regardless of their probability, which significantly increases our sample size, but only slightly increases the photometric redshift uncertainty. The fraction of catastrophic outliers for the full sample is $\sim 6\%$.  

If we consider only galaxies in the redshift range of $1.5 < z < 3$ (as we do in this study; see Section \ref{sec:complete}), we find that 133 of the 1282 members of the final catalog have spectroscopic redshifts.  We find an average offset $\langle \Delta z/(1+z) \rangle = 0.06$ and a rms of $\sigma_{\Delta z} = 0.10$, with a fraction of catastrophic outliers of $20\%$.  

The most reliable photometric redshifts in the $z>1$ regime have so far come from studies of galaxies that have extensive multi-band coverage \citep{Ilbert2009,Cardamone2010}.  For example, \citet{Ilbert2009} derive photometric redshifts for the COSMOS sample of galaxies out to $z\sim1.25$ using 30 unique filters and accounting for the emission line contribution to broad-band magnitudes.  They find an accuracy of $\langle \Delta z/(1+z) \rangle = 0.007$.  Using the same technique for the zCOSMOS galaxies at $1.5<z<3$, \citet{Ilbert2009} find an accuracy of $\langle \Delta z/(1+z) \rangle = 0.06$ and a 20\% fraction of catastrophic failures.  This is similar to the range found in this study, but note that they define catastrophic outliers as objects with accuracies worse than 15\% (as opposed to the 30\% used here).  \citet{Cardamone2010} use 18 medium-band filters for a sample of galaxies between $1.2 <z< 3.7$.   They determine an accuracy of photometric redshifts of $\langle \Delta z/(1+z) \rangle = 0.02$ and a rms of $\sigma_{\Delta z} = 0.03$.   

We caution in the paper that the reliability of our photometric errors is not as high as with studies that are able to use 18 to $>30$ filters to determine photometric redshifts.  However we note that when comparing our photometric redshift accuracies with a study that use a similar number of filters to derive photometric redshifts, we find that the reliability is in good agreement.   Using the CDFS, \citet{Damen2009}  find a median $\langle \Delta z/(1+z) \rangle = 0.03$ for their full sample and $\langle \Delta z/(1+z) \rangle = 0.079$, for galaxies with $z>1$.   Again we stress that the unique aspect of this study is ability to measure stellar masses at a deeper level than previously given our deep H-band HST data.   Redshifts are a critical aspect of this study however, and we revisit this issue in detailed in Section~\ref{mc} where we discuss the possible effects of the photometric redshift errors on all our results through several simulations.

\subsection{Stellar Masses and Restframe Colour}\label{mass}

Stellar masses and restframe colours are determined from multicolour stellar population fitting techniques using the same catalog of five broad band data points used to determine photometric redshifts for all GNS galaxies.  A detailed description of how stellar masses and rest-frame $(U-B)$ colours are derived can be found in \citet{Conselice2011} and \citet{Grutzbauch2011}, and is summarised in the following.  

To calculate these masses and colours we construct a grid of model spectral energy distributions (SEDs) from \citet{BC03} stellar population synthesis models, assuming a Salpeter initial mass function and a varying star formation history, age, metallicity, and dust extinction.  The star formation history is characterized by an exponentially declining model of the form $\psi (t) = \psi _{\circ} \exp(-t/\tau)$ with the star formation timescale ranging from $\tau$~=~0.01 to 10 Gyr and the age of the onset of star formation ranging from $0$ to $10$ Gyr.    The dust content is parametrized by the $V$-band optical depth with values of $\tau _{V} = 0.0, 0.5, 1, 2$, and the metallicity ranges from 0.0001 to 0.05 \citep{BC03}.  

The magnitudes resulting from the model SEDs are then fit to the observed photometric data of each galaxy using a Bayesian approach.  For each galaxy, a likelihood distribution is computed for stellar mass, age, and absolute magnitude for each possible star formation history.  We chose to compute rest-frame $(U-B)$ colours, since the wavelength range of the $U$- and $B$-bands are covered best by the observed optical and $H$-bands.  The peak of the likelihood distribution is adopted as the galaxy's stellar mass and $M_{U}$ and $M_{B}$ absolute magnitudes, while the uncertainty of these values is determined by the width of the distribution.  While parameters such as age, e-folding time, dust extinction, and metallicity are not accurately fit due to various degeneracies, the stellar masses and colours are robust. From the width of the probability distributions we determine typical errors for our stellar masses of 0.2 dex. There are additional uncertainties from the choice of the IMF and due to photometric errors, resulting in a total random error of our stellar masses of $\sim$0.3 dex, roughly a factor of two.  The rest-frame $(U-B)$ colours  have errors of  $\sim0.2$.

\subsection{Sample Selection and Completeness}\label{sec:complete}

\begin{figure}
\includegraphics[width = 0.45\textwidth]{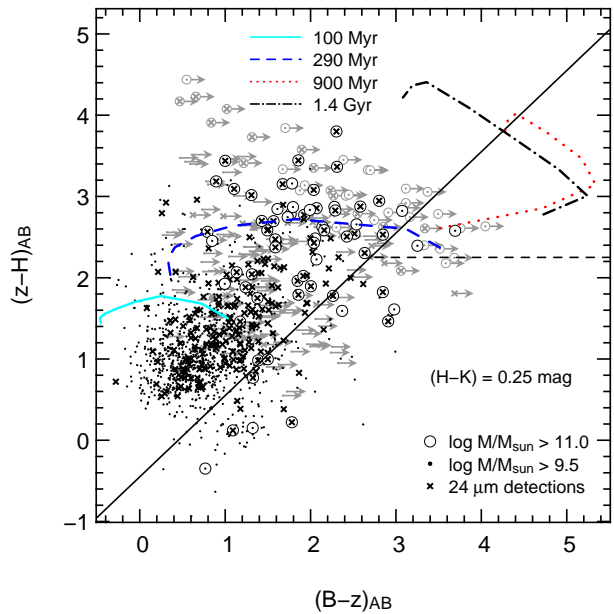}
\caption{
$(z-H)$ vs. $(B-z)$ diagram for GNS galaxies with log~$M_{*}\ge9.5$ and $1.5<z<3.0$.  Large circles show galaxies with log~$M_{*}\ge11$ and crosses identify galaxies also detected at 24 $\mu$m. The solid diagonal line and dashed horizontal line represent the constraints of criteria of \citet{Daddi2004} for star-forming galaxies.  The other lines represent the evolution of single stellar populations of given ages as observed at $1.5<z<3.0$.  Galaxies with ACS/$B$-band magnitudes fainter than 28.2 (AB) are given this value as a limit and shown in grey with arrows.   \label{bzh}    }
\end{figure}

For this work, we use a complete sample of galaxies selected by stellar mass.  In this way we can perform an analysis of the assembly of all galaxies down to log~$M_{*}= 9.5$ at high redshift without obvious biases towards star-forming galaxies.  A detailed analysis of stellar mass function for the GNS Survey is presented in \citet{Mortlock2011}.  Here we summarise the method of selection by stellar mass.   

To determine the limiting stellar mass for the GNS, we compute the expected completeness limits from the 5 $\sigma$ magnitude limit of the survey and the mass-to-light ratios ($M/L$) of simple stellar populations (SSP).  The $M/L$ ratio at the rest-frame wavelength corresponding to the observed NICMOS/$H$-band at $z \sim 3$ is determined for a maximally old galaxy using the models of \citet{worthey1994}.  The $M/L$ ratio is then multiplied with the luminosity corresponding to a $z = 3$ galaxy at the detection limit of $H_{AB} = 26.8$ to obtain the limiting stellar mass.  In this way, we calculate a stellar mass completeness limit of log~$M_{*}=9.2$ at $z=2.75$ \citet{Mortlock2011}.  

The deep limiting $H$-band magnitude easily exceeds that of the expected upper bound for dusty bright sub-millimeter galaxies of 23.3 magnitudes found by \citet{Frayer2004}, so we are confident that we are not missing the dustiest galaxies due to attenuation.  But even so, to test if we are biased against low mass dusty star-forming objects we examine with stellar population models if we would find, for example, a very low mass dusty galaxy at $z=2.5$.  If this object is composed of 50\% old stars and 50\% 100~Myr stellar population, this object would be 2.82 magnitudes brighter in M$_{B}$ than a galaxy composed of a purely old stellar population formed at $z=10$.  Thus, for this object to be undetected at the very faint end of our completeness would require a dust extinction of 3.16 magnitudes, which is an amount of dust extinction exceeding what is commonly found in the lowest mass galaxies in our sample.  That is, we would be detecting these extremely dusty low mass galaxies at our depth, but are not likely because they do not exist in any great numbers.     

From the original GNS catalog of 8298 galaxies, we select a sample of galaxies complete down to log~$M_{*}= 9.5$ and $1.5<z<3.0$, which corresponds to a sample size of 1282 galaxies within a survey volume of  $2.3 \times 10^{5}~$Mpc$^{3}$.  We find that 8\% of the galaxies are massive (log~$M_{*} \ge 11$) and 10\% have spectroscopic redshifts.  Our final sample is the largest HST-based dataset with rest-frame optical imaging of mass-selected galaxies complete down to log~$M_{*}=9.5$ over $1.5<z<3$.  

To gain a preliminary understanding of the properties of this mass-selected sample, we look at the optical and near-infrared two-colour plot developed by \citet{Daddi2004} as a technique to determine whether galaxies at $z\sim2$ show properties of star-forming or passive systems (the $BzK$ method).  With the assumption that $(H-K) = 0.25$ on average at our redshifts \citep{brammer2007}, we show the corresponding $BzH$ colour-colour space in Figure~\ref{bzh} for our complete sample.  Black points show galaxies with 5~$\sigma$ detections at all three magnitudes.  The full mass-complete sample was selected from H-band observations to $H \le 26.8$.  The $z_{850}$-band depth is $z = 27.5$ which results in a 98.5\% (1263/1282) detection rate among the full mass-selected sample.  The 5~$\sigma$ detection limit in the $B_{435}$-band is $B=28.2$ which detects 87.5\% (1120/1282) of the galaxies from the mass-selected sample.  

Non-detections are assigned the limiting value ($B$ = 28.2) and are shown as grey arrows in Figure~\ref{bzh}.  Small points show all GNS galaxies with log~$M_{*}\ge 9.5$ while large circles represent massive galaxies, with $M_{*}\ge10^{11}\Msun$.  Crosses represent galaxies also detected at 24 $\mu$m with $f_{24}>30\mu$Jy.  The solid diagonal black line and dashed horizontal black line show the $BzH$ regions defined by $(z-H)~-~(B-z) < -0.2 - (H-K)_{AB}$ and $(z-H) > 2.5 - (H-K)_{AB}$.  The various tracks show passively evolving \citet{BC03} stellar population models as they would be observed between $z=1.5-3$ from a single burst occurring at the ages indicated at the upper left.

Most of the GNS galaxies in Figure~\ref{bzh} sit to the upper-left of the solid diagonal line, indicating they are star-forming galaxies according to the criteria of \citet{Daddi2004}.  More massive galaxies tend to have redder $(z-H)$ and $(B-z)$ colours, with almost all the massive galaxies showing $(z-H)>2$.  Galaxies detected at 24~$\mu$m are spread throughout the colour-colour space.  In contrast, there are only two galaxies (detected in all three bands) within the triangular region to the upper right, indicative of passively evolving stellar populations.  It is notable that neither of these galaxies are detected at 24 $\mu$m.  Among the galaxies with limiting $B$-band magnitudes (grey) in the passive region, only one is detected at 24~$\mu$m.  The dot and dash-dot evolutionary tracks in the passive region of Figure~\ref{bzh} show galaxies of age 900 Myr and 1.4 Gyr, respectively, as they would appear between $z=3$ and $z=1.5$ when the universe was between 2.2 and 4.4 Gyr old.  This shows that using colour-colour diagrams does not work well for selecting red galaxies since you need extremely deep $B$-band observations.

\subsection{Supplementary Infrared Data}\label{sec:irdata}
The Spitzer GOODS Legacy program \citep{Dickinson2003} provides deep infrared observations covering the GOODS fields, which we use to calculate infrared properties for GNS galaxies.  Since the resolution of the Spitzer Multiband Imaging Photometer (MIPS) 24 $\mu$m images is $6 ^{\prime\prime}$ FWHM, we use the Spitzer Infrared Array Camera (IRAC) data to perform source matching within a radius of $1^{\prime\prime}$. IRAC data were reduced with the pipeline version S18.7.0 and then mosaicked with {\tt MOPEX}, a package developed at the Spitzer Science Center for image processing.  As described in Appendix A in \citet{pg_2008_z4}, IRAC sources detected with {\tt SExtractor} (down to a 5-$\sigma$ level of $f(3.6\mu m)\sim24.0$~mag) were cross-correlated with UV-to-NIR images and photometry was measured in consistent apertures to build complete SEDs.  A complete list of the extended data sets used, and their properties, is given in \citet{pg_2008_z4}.  For the work presented in this paper, the redshifts were fixed to the values used throughout this paper and in \citet{Grutzbauch2011}.

The photometric method also includes a deblending algorithm for the IRAC and MIPS images which takes the known positions of the blended sources obtained from optical ACS images and the point spread functions (PSFs) for the different images, and obtains separated fluxes for the blended sources (see also \citet{Grazian2006}). 

This paper makes use of the MIPS observations of GNS galaxies to calculate infrared luminosities and IR star formation rates. The MIPS 24~$\mu$m data are those obtained as part of the GOODS Guaranteed Time observations and the Far-Infrared Deep Extragalactic Legacy Survey (FIDEL), of both North and South fields. The final mosaic has a 5-$\sigma$ detection level of 30~$\mu$Jy within the GOODS area. Catalogs were made with a PSF-fitting method described in \citet{PerezGonzalez05,pg_2008_z4}. The MIPS catalog has an 80\% completeness limiting flux of 30 $\mu$Jy for the GOODS fields, based on simulations aiming to recover artificial sources inserted into the images \citep{Papovich2004}.  

One potential problem with the above matching procedure is that there could be some instances where a given IRAC source could be matched to several GNS galaxies.  The deblending procedure we use  accounts for this problem in most cases.  However, we have identified seven instances where one infrared detection could be associated with multiple HST/NICMOS sources.  We have removed these seven objects from this analysis so that no ambiguity remains.  This means that all galaxies in the mass complete GNS sample of 1282 galaxies have IRAC counterparts that are secure.  MIPS detections are not as high, with $\sim$21\%~(267/1282) of the GNS galaxies detected above a flux limit of f$_{24}=30~\mu$Jy.  

The total infrared luminosity, $L_{IR}$ (8-1000$\mu$m), is determined following the procedure described in \citet{pg_2008_spitz} and \citet{pg_2008_z4}.   Using the observed 24 $\mu$m flux density, we use the SED templates of \citet{CE2001} to best reproduce the corresponding rest-frame luminosity at $\sim8\mu$m at $z=1.5-3$, and then obtain the integrated $L_{IR}$ of that template.  If a source is also detected at 70 $\mu$m, then a colour is used in the comparison with templates, instead of a monochromatic luminosity.  We use the $L_{IR}$ in Section~\ref{sec:lir} to determine infrared SFRs.  Rest-frame luminosities at 1600$\mathrm{\AA}$ and 2800$\mathrm{\AA}$ were also estimated from the interpolation of the best-fitting templates at these respective wavelengths.  A UV-slope was then calculated from the interpolated monochromatic luminosities, which will be further discussed in Section~\ref{sec:dust}.

\section[]{Star Formation Activity}\label{sec:sfrall}

A major goal of this paper is to determine, for the first time, the star formation activity present in a stellar-mass-selected sample of galaxies in the redshift regime $1.5~<~z~<~3$.   Determining star formation activity at these redshifts is not trivial.  Infrared observations are useful indicators of dust heating due to star formation, but current Spitzer Space Telescope observations are not yet deep enough to detect the full mass-selected sample of galaxies, as only 60\% of the massive galaxies are detected, and only $\sim$21\%~(267/1282) of the entire sample are detected at 24 $\mu$m (above a flux limit of f$_{24}=30~\mu$Jy; see Section~\ref{sec:irdata}). 

On the other hand, 98.5\% of the mass-selected GNS sample are detected in the rest-frame ultraviolet (observed ACS/$z_{850}$-band), which is sensitive to young ($<1$ Gyr old) stars even for high redshift, optically red galaxies.   \citet{Kaviraj2008} conclude that in their sample of early-type galaxies at $0.5<z<1$, the UV flux comes exclusively from young stars and the same holds for our even younger galaxy sample.  Furthermore, they find that only $\sim$1.1 percent of those galaxies are consistent with purely passive ageing since $z=2$, providing confidence that the rest-frame UV flux from galaxies in this study is highly sensitive to young stars, and therefore representative of recent star formation.  In this section, we derive and compare star formation rates from both rest-frame ultraviolet and infrared wavelength regimes and determine the best star formation measure to use for the rest of this study.

\subsection{Ultraviolet Star Formation Rates}\label{sec:sfr}

We use rest-frame UV light as a SFR indicator, which we measure from the observed optical ACS/$z_{850}$-band.  The UV luminosity is closely related to the level of ongoing star formation as the UV continuum luminosity is produced by short-lived O and B stars, even at redshifts of $z>1$ \citep{Kaviraj2008}.  We determine the SFR$_{UV}$ from the observed optical ACS/$z_{850}$-band flux density (with a 5~$\sigma$ limit of 27.5 in the AB system), which corresponds to the rest-frame UV luminosity around 2800$\mathrm{\AA}$, spanning wavelengths of 2125~-~3400$\mathrm{\AA}$ for $z=1.5-3$ galaxies.  Of the 1282 GNS galaxies in our sample with log~$M_{*}\ge9.5$ and $1.5<z<3$, only 1.5\% do not have counterparts in the ACS/$z_{850}$-band catalog (see Section~\ref{sec:gns}).  

To calculate SFR$_{UV}$, first we derive absolute magnitudes using the {\tt kcorrect} package \cite[v4.2]{Blanton2007}.  We then use the \citet{Kenn98} conversion from 2800$\mathrm{\AA}$ luminosity to SFR assuming a Salpeter IMF: 

\begin{equation} \label{eq:sfr}
\textrm{SFR}_{\mathrm{UV}}\,(\Msun \,\textrm{yr}^{-1}) = 1.4 \times10^{-28}\, L_{2800}\,(\textrm{ergs }\,\textrm{s}^{-1}\,\textrm{Hz}^{-1}).
\end{equation}
Before dust extinction is taken into account, we find at $z~=~1.5$ a limiting SFR$_{UV,obs}= 0.28\pm0.1~\Msun$yr$^{-1}$ and at $z~=~3.0$, we find a limit of SFR$_{UV,obs}= 0.98\pm0.3~ \Msun$yr$^{-1}$.  The errors quoted here take into account photometric errors and the error in the conversion from a luminosity.  The overall errors for individual SFRs are about 30\%, which are dominated by the dust correction applied, as we discuss in the next section.

\subsection{Dust Absorption Determination}\label{sec:dust}

As is now well known, SFR estimates derived from UV light can be highly affected by dust extinction.   Inferring the level of dust reddening from photometric properties remains a major challenge for understanding the star-forming properties of distant galaxies.  To address this issue, we examine techniques to recover UV light absorbed by dust and re-radiated at longer wavelengths.  

\begin{figure}
\includegraphics[width = 0.47\textwidth]{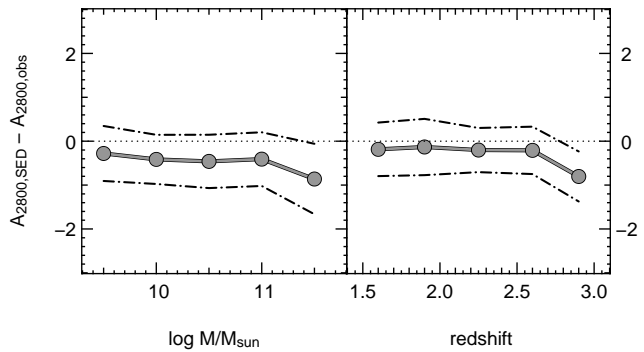}
\caption{\label{Av_compare}
The difference between A$_{2800,SED}$ based on an SED-determined UV slope and A$_{2800,obs}$, derived from the observed $(B-z)$ colour.     }
\end{figure}

\begin{figure*}
\includegraphics[width = 0.8\textwidth]{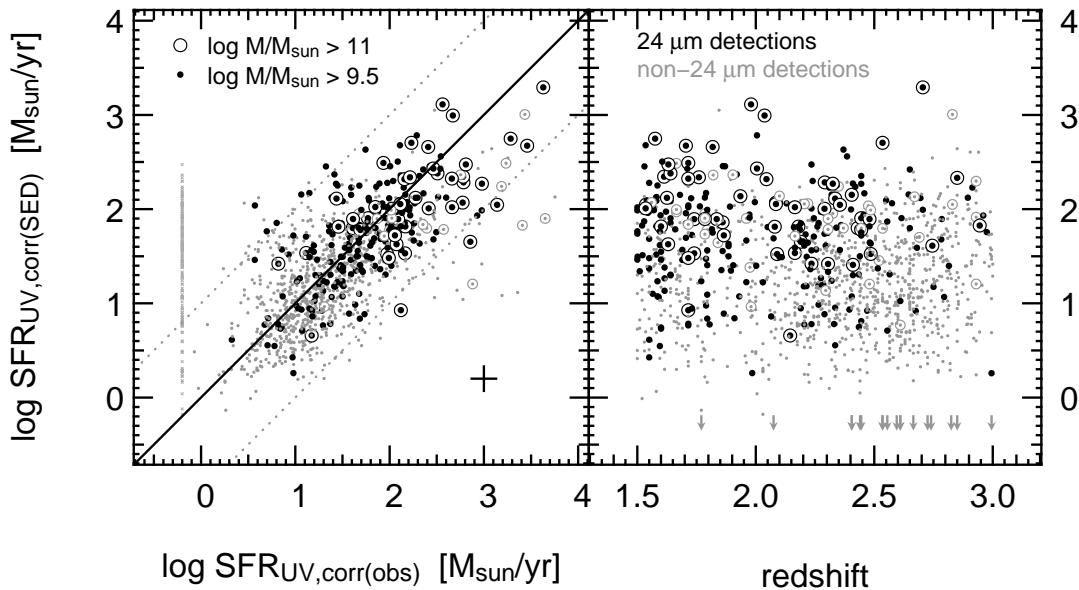}
\caption{\label{sfruv_compare}
The left panel compares SFR$_{UV}$ derived from observed $z_{850}$-band with different dust corrections applied.   Along the X-axis of the left-hand panel is SFR$_{UV,corr(obs)}$, corrected by observed optical colours; along the Y-axis is SFR$_{UV,corr(SED)}$, corrected from an SED-determined UV slope.   The solid line shows the one-to-one relation while the dotted lines shows factors of 10 difference.  The right panel shows SFR$_{UV,corr(SED)}$ as a function of redshift.  Open circles identify massive galaxies with log~$M_{*}\ge11$.  Small grey points show all GNS galaxies with log~$M_{*} \ge9.5$ and $1.5<z<3$ not detected at 24 $\mu$m.   Black points show galaxies detected at 24 $\mu$m with $f_{24} >$ 30 $\mu$Jy.  Grey arrows show the redshifts of objects fainter than the 5~$\sigma$ $z_{850}$-band detection limit of 27.5 mag (AB).  Points in the left plot at SFR$_{UV,corr(obs)}=-0.2$ show galaxies below the 5 $\sigma$ $B$-band limit.  
 }
\end{figure*}

One method for correcting for dust obscuration is to use the UV slope as an estimate for the amount of extinction.  \citet{Meurer1999} demonstrated a correlation between dust attenuation and rest-frame UV slope, $\beta$, for a sample of nearby starburst galaxies (where $F_\lambda \sim \lambda^{\beta}$). Updated studies of local galaxies using the {\it Galaxy Evolution Explorer} (GALEX) near-ultraviolet band \citep{buat2005,seibert2005} and $z\sim2$ galaxies \citep{Reddy2010} show that the UV slope from the local starburst relation can be used to recover the dust attenuation of a vast majority of moderately luminous galaxies at $z\sim2$, but with a large scatter of $\sigma \sim 0.9$ mag \citep{seibert2005}.   Dust attenuation measurements derived from the UV for  galaxies at $z\sim2$ have been examined for samples of galaxies selected from several different methods: the $BzK$ method \citep{Daddi2004,Pannella2009}; Lyman Break Galaxies (LBGs) \citep{Reddy2010}; and rest-frame UV \citep{Erb06-ha}.  However, as is well known for red galaxies, sorting out the difference between dusty galaxies and those with evolved stellar populations can be difficult \citep[e.g.][]{vandokkum2010,Viero2010}.  

Another method for determining dust extinction relies on using the reddening parameter value extracted from a best-fitting SED template.   As described in Sections~\ref{mass} and~\ref{sec:irdata}, we fit sets of templates built from the stellar population synthesis models twice for this sample:  once to derive the stellar masses \citep{Grutzbauch2011,Conselice2011} and separately to determine the infrared lumniosity \citep{pg_2008_spitz}.  

Unfortunately, neither of our SED fits provide reliable values of dust extinction and we therefore rely on the UV slope method to determine a dust attenuation, A$_{2800}$.  There are some notable limitations and caveats that exist when using this method to correct for dust, since the technique assumes that the UV slope is due to dust reddening, rather than other sources such as evolved stellar populations.  Despite the difficulties, we are encouraged by the facts that a large majority of the GNS sample shows colours of galaxies experiencing some level of star formation as in Figure~\ref{bzh} and 98.5\% are detected in the rest-frame UV.  

Here we apply two methods for determining a UV dust attenuation, A$_{2800}$, in terms of the UV slope, as well as investigate properties of our red distant galaxies (see also \citet{Grutzbauch2011} for a discussion of this).  But considering the difficulties in determining dust extinctions in the UV, for the results presented in this paper, we show SFR$_{UV}$ before and after correcting for dust.  

The first method we investigate is an empirical technique for star-forming galaxies that comes from observed optical data.   \citet{Daddi2004} calibrated the correlation between $E(B-V)$ and the observed optical $(B-z)$ colour using a BzK-selected sample of star-forming galaxies with $K<$ 20 in the redshift range $1.4<z<2.5$.  Combining this with the \citet{Calzetti00} reddening law, this becomes:

\begin{equation} \label{eq:dust}
A_{2800,obs} = 1.815\ (B-z+0.1)_{\textrm{AB}}.  
\end{equation}
We note that this method is only applicable to the 87.5\% of the mass-selected GNS galaxies that are detected above the 5$\sigma$ $B_{AB}$ limit of 28.2.   Applying this method we find an average value of A$_{2800,obs}$ of 1.9~$\pm$~1.2 mag.

The second method we use to determine A$_{2800}$ utilises the SED-fitting procedure described in Section~\ref{sec:sfr}.  Since we do not recover a reliable A$_{V}$ as a free parameter of the fit, we use the best-fitting SED to extrapolate and measure UV luminosities at wavelengths of 1600$\mathrm{\AA}$ and 2800$\mathrm{\AA}$, and use these to calculate the UV slope $\beta$.  This should be a more reliable approach to uncovering the dust extinction as it includes older evolved SEDs, and provides a value of A$_{2800}$ for every galaxy.  

We follow the \citet{Calzetti00} methodology to derive A$_{1600}= 2.31\beta + 4.85$ and A$_{2800,SED}$ as A$_{2800,SED} =$A$_{1600} (k_{2800} / k_{1600})$.  The slope, $\beta$, is calibrated for the values $-2.0<\beta<1.0$ with the \citet{Calzetti00} law and for this reason, if the SED-derived $\beta$ is less than $-2.0$ (144 objects), we assign a value of A$_{2800} = 0.0$.  If $\beta > 1.0$ (60 objects), we assign an A$_{2800} = 3.5$ so that there is not a large overcorrection for dust.  Both of these cases are however rare.   A similar upper limit of A$_{FUV} \sim3$ was employed by \citet{Salim2007} based on the $(FUV-NUV)$ colour observed by GALEX.  Using this method we find an average value of A$_{2800,SED}=1.6~\pm$~1.2 mag.  \citet{Tresse07} study galaxies from the VIMOS VLT Deep Survey (VVDS) and find similar values for far-UV dust obscuration of $1.8-2$ mag from $z=2$ to $z=0.4$.

Figure~\ref{Av_compare} compares the two separate determinations of A$_{2800}$ as a function of stellar mass and redshift.  They are in relatively good agreement for $M_{*}<10^{11}\Msun$ with the absolute value of $<\Delta A>=0.3$ and standard deviation of 0.6.  The scatter increases for $M_{*}\ge10^{11}\Msun$ galaxies showing that there is a difference between the empirical and SED approach towards dust corrections.  Values of A$_{2800,obs}$ tend to be larger than A$_{2800,SED}$ for $M_{*}\ge10^{11}\Msun$ galaxies by the amount $\delta$A$_{2800}= 0.86\pm0.7$.  As shown in the right panel of Figure~\ref{Av_compare}, the two measures of A$_{2800}$ are, however, in rough average agreement across the entire redshift range, with $<~\Delta ~A>~=~0.18\pm0.03$ for $1.5<z<2.6$.  At $z>2.6$, the A$_{2800,obs}$ begins to leave the redshift range for which it was calibrated and we see a systematic offset such that A$_{2800,obs}$ is larger than A$_{2800,SED}$ by a value of 0.8.  These results confirm recent findings by \citet{Nordon2010} who stack Herschel observations from the PACS instrument to show that  \citet{Daddi2004} calibration overestimates the dust attenuation.  

In Figure~\ref{sfruv_compare}, we show how these separate dust corrections affect the actual SFR measures.  The Y-axis shows the SFR$_{UV,corr(obs)}$ corrected with the empirical method (A$_{2800,obs}$) and along the X-axis we show SFR$_{UV,corr(SED)}$, corrected using the SED fitting based dust correction (A$_{2800,SED}$).  Nearly all measurements of these two dust corrections agree within a factor of 10 (grey dotted lines).   

There is evidence for an offset present among massive galaxies not detected at 24 $\mu$m (open grey circles with dots) in Figure~\ref{sfruv_compare}. This difference is such that the dust correction derived from the observed UV slope, and thus the resulting star formation rate, is larger than the SFR corrected with the SED-derived UV slope.  This difference can be quite a bit larger, over a factor of ten. This demonstrates that the SED method for finding the dust extinction, which we ultimately use, is able to locate those systems which are not dusty, but red due to older stellar populations.  The star formation rates corrected with the SED-derived dust extinction are also on average lower than the empirically corrected SFRs, and are thus if anything underestimating the star formation rates for these galaxies.   

What is clear from this comparison is that there are red galaxies, particularly with large masses, that have older stellar populations, creating their red colours.   The A$_{2800}$ values calculated from the observed $(B-z)$ colour was originally derived by \citet{Daddi2004} from a set of galaxies chosen to be star-forming, and therefore will not account for galaxies with old stellar populations.  Since the determination of the UV slope from observed passbands was originally derived from star-forming galaxies only, we apply the SED-derived dust correction to our observed SFR$_{UV,obs}$ and use this as the main SFR indicator throughout the rest of the paper.  

Furthermore, the validity of our dust corrections are backed up by comparing  the fraction of galaxies detected at 24$\mu$m (\S~\ref{sec:lir}) as a function of our completely independently derived values of $\beta$.  The left panel of Figure~\ref{f24} shows the fraction of galaxies detected at 24 $\mu$m as a function of $\beta$.  We find a step decline in the fraction of galaxies detected from $\sim 60$\% at $\beta \sim 1.5$ down to 15\% at $\beta \sim -1.5$.  This shows that our measures of dust extinction, through the value of $\beta$ derived from SEDs, correlates well with the presence of 24$\mu$m emission from heated dust, as would be expected.  

The dust corrected SFR$_{UV,corr}$ that we will use throughout the rest of this paper (the rest-frame UV SFR derived from the observed ACS/$z_{850}$-band, corrected using the SED-determined A$_{2800}$ values) is shown in the right panel of Figure~\ref{sfruv_compare} as a function of redshift.  We find that our galaxies exhibit a range in SFR$_{UV,corr}$ between 0.6 $\Msun$yr$^{-1}$ and $\sim$2000 $\Msun$yr$^{-1}$.  The average SFR$_{UV,corr}$ for the full mass-selected sample between $1.5<z<3$ is $42 ^{+53}_{-29}~\Msun$yr$^{-1}$.  For massive galaxies with $M_{*}\ge10^{11}\Msun$, we find an average SFR$_{UV,corr}$ of $103\pm75~\Msun$yr$^{-1}$ (consistent across the full redshift range).  Among galaxies detected at 24 $\mu$m ($\sim$21\% of the sample), we find average SFR$_{UV,corr}=73\pm50~\Msun$yr$^{-1}$.  The errors quoted are the dispersion of the values of SFR in the sample.  The majority of SFRs of individual galaxies have errors of 0.2 dex, which are dominated by the dust correction applied. Furthermore, our values of  A$_{2800}$ and SFR$_{UV}$ are consistent with those derived by \citet{Onodera2010} who derive SFR$_{UV}$ from rest-frame 1500$\mathrm{\AA}$ luminosity.  They observe a sample of star-forming BzK galaxies at $z\sim2$ and determine dust corrections from optical emission lines and test their agreement with the rest-frame UV slope.  Overall our dust-corrected values of SFR$_{UV}$ within their stellar mass range of $M_{*}>5\times10^{10}\Msun$ are consistent with their findings.

\subsection{Infrared Star Formation Rates}\label{sec:lir}

\begin{figure}
\includegraphics[width = 0.5\textwidth]{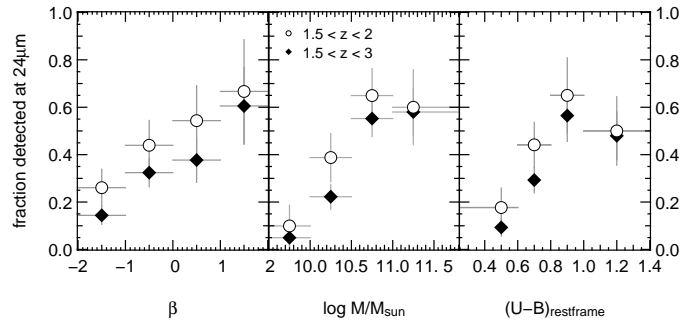}
\caption{\label{f24}
Fraction of galaxies detected above the limiting 24 $\mu$m flux of 30 $\mu$Jy as a function of UV slope $\beta$, stellar mass, and rest-frame $(U-B)$ colour.  Fractions derived from the full redshift range are shown as solid diamonds, while galaxies restricted to the lower redshift range of $1.5<z<2$ are shown as open circles.  }
\end{figure}

Star formation rate and dust properties of distant galaxies can also be measured through emission from the infrared.  This light originates from dust in galaxies which has been heated by photons from massive stars, and thus in principle should agree with measurements of the star formation rate derived directly from these stars in UV light once corrected for dust extinction.  

Over the full redshift range from $1.5<z<3$, we find that $\sim$21\%~(267/1282) of our sample is detected at 24 $\mu$m, above a flux limit of f$_{24}=30~\mu$Jy.   If we restrict the redshift range to $1.5<z<2$ then we find a higher fraction of 35\% detected.  In Figure~\ref{f24}, we show detection fractions at 24~$\mu$m for galaxies over the full redshift range (solid diamonds) and for a subsample of galaxies restricted to a lower redshift range of $1.5<z<2$ (open circles) separately.  The trends are qualitatively similar for both groups and we give values for the full redshift range unless otherwise indicated.   We find that the detection fraction at 24 $\mu$m depends on stellar mass and rest-frame $(U-B)$ colour.  Nearly 60\% of the GNS galaxies with log~$M_{*} > 10.8$  are detected at 24 $\mu$m.  The detection fraction decreases to $\sim$20\% at log~$M_{*}\sim10.25$ and less than 5\% for log~$M_{*} < 10$ galaxies.    Between 50 and 60\% of galaxies with $(U-B) \ge 0.8$ are detected at 24 $\mu$m while the detection fraction decreases at bluer colours.  A detection fraction of 30\% is found for galaxies with $0.6<(U-B)<0.8$ and 10\% for the bluest galaxies at $(U-B)<0.6$.  This implies that more massive galaxies and galaxies with red colours have higher obscured SFRs and more dust, causing enough emission at 24~$\mu$m to be detected.  

\begin{figure*}
\includegraphics[width = 0.7\textwidth]{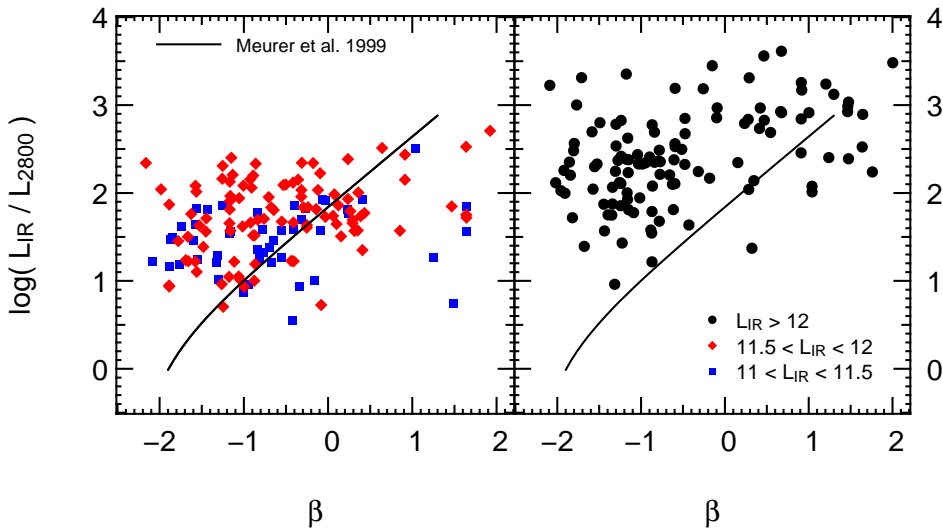}
\caption{\label{irx_beta}
The ratio of $L_{IR}$(8-1000$\mu$m) to $L_{UV}$(2800$\mathrm{\AA}$) as a function of the UV-slope, $\beta$, for $1.5<z<3.0$ galaxies with f$_{24}>30 \mu$Jy.  The solid line shows the local relation from \citet{Meurer1999}.  Black points in the right panel show galaxies classified as ULIRGs (log~$L_{IR}>12$).  In the left panel, galaxies with $11.5<$~log~$L_{IR}<12$ are shown as red diamonds and galaxies with $11<$~log~$L_{IR}<11.5$ are shown as blue squares. }
\end{figure*}

We can get some idea about the relative SFRs measured from UV and FIR light through Figure~\ref{irx_beta} which presents the ratio of $L_{IR}$ to $L_{2800}$, as a function of the SED-derived UV slope, $\beta$.  Galaxies with lower infrared luminosities, $11<$~log~$L_{IR}<11.5$, most closely mimic the local relation for starburst galaxies derived by \citet{Meurer1999}.  As infrared luminosity increases, the $L_{IR}$ to $L_{2800}$ ratio increases such that nearly all ULIRGs (with log~$L_{IR}\ge12$) lie above the \citet{Meurer1999} relation (right side of Figure~\ref{irx_beta}).  

In our sample of galaxies, we find that 44\% of all galaxies detected at 24 $\mu$m are classified as ULIRGs.  These results largely agree with the $z\sim2$ star-forming galaxies studied in \citet{Reddy2010} who select UV-bright Lyman break galaxies.  Our values have a larger spread than \citet{Reddy2010}, most likely due to the fact that we are utilising a mass-selected sample.  This larger variation of star-forming properties may be especially prominent among massive galaxies \citep[e.g.][]{vandokkum2010}.  Similarly, using the GOALS sample \citet{Howell2010} find that LIRGs have larger IR excesses than lower luminosity galaxies of similar UV colour. On average, we also find that LIRGs and ULIRGs have larger IR excesses.

\begin{figure}
\includegraphics[width = 0.45\textwidth]{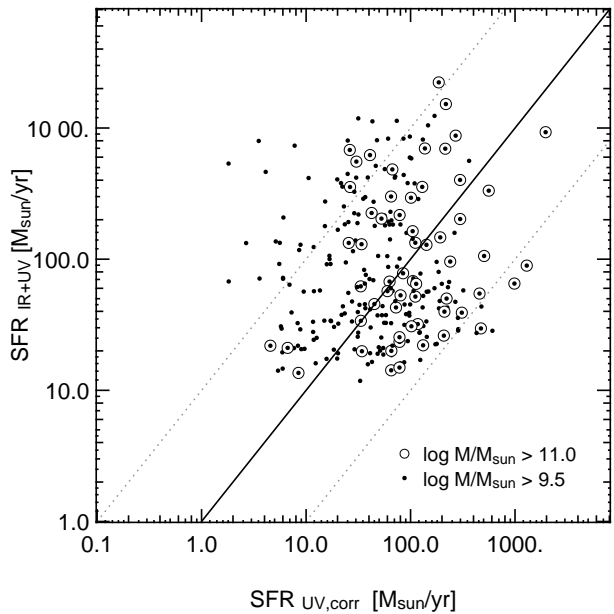}
\caption{\label{sfr_uvIR}
SFR$_{IR+UV}$ versus SFR$_{UV,corr}$ for all galaxies detected at 24 $\mu$m.  Open circles identify massive galaxies with $M_{*} > 10^{11} \Msun$.   The solid line shows the one-to-one relation while the dotted lines shows factors of 10 difference.  }
\end{figure}

To derive infrared SFRs from from $L_{IR}$, we use the \citet{Kenn98} conversion,

\begin{equation} \label{eq:sfr}
\textrm{SFR}_{IR}\,(\Msun \,\textrm{yr}^{-1}) = 4.5 \times10^{-44}\, L_{IR}\,(\textrm{ergs }\,\textrm{s}^{-1}).
\end{equation}
We then determine the total SFR in the infrared for each galaxy by adding the SFR$_{IR}$ together with the observed value of SFR$_{UV,uncorr}$.    Figure~\ref{sfr_uvIR} directly compares SFR$_{IR+UV}$ and SFR$_{UV,corr}$ for mass-selected GNS galaxies.  Among massive galaxies (open circles), $\sim$87\% agree in SFR within a factor of 10, while 80\% of all the galaxies in Figure~\ref{sfr_uvIR} agree to within a factor of 10.  However, there is no real correlation between these measures and the scatter is large.  

We find that the SFR$_{IR+UV}$ is an average factor of 3.5~$^{+4.6}_{-2.1}$ larger than SFR$_{UV,corr}$ between $1.5<z<3$, an offset which has been seen before in galaxies at high redshift \citep[e.g.][]{Papovich2007,Elbaz2010}.  There are several factors that could contribute to this offset.  A recent study by \citet{Elbaz2010} compares Herschel deep extragalactic surveys from 100 to 500 $\mu$m  total IR emission extrapolated from 24 $\mu$m Spitzer fluxes.  They estimate that at $z>1.5$, the 24 $\mu$m observations overestimate the SFR by a factor of $\sim$3, which could be due to a larger PAH emission in distant galaxies, hot dust heated by a buried AGN, or uncertainty in local SED templates.  

Some of these issues are explored in a forthcoming study by \citet{Weinzirl2011} of massive GNS galaxies with $M_{*} > 5\times10^{10} \Msun$ in the redshift range $1<z<3$.   They determine SFR$_{IR}$ for their sample independently of this study using infrared luminosities derived from 24~$\mu$m data and apply the factor of $\sim$3 correction suggested by the early results from Herschel \citep[e.g.][]{Elbaz2010,Nordon2010}.  For massive galaxies, they find that the SFR$_{IR}$ agrees with the SFR$_{UV,corr}$ values within a factor of $\sim$ 2 to 3 for galaxies with SFR$_{IR} > 40 \Msun$yr$^{-1}$, but galaxies with low SFR$_{IR}$ are lower than SFR$_{UV,corr}$ by up to a factor of 10.
  
It is also possible that the UV-based SFRs could be underestimated in cases of very high extinction, where star formation could be completely unrecoverable by UV observations alone.  It is unlikely, however, that we are missing many star-forming galaxies altogether from our samples though, since the primary selection is done from deep $H$-band imaging and a high level of extinction would be needed for an object to be completely undetectable \citep[see Section~\ref{sec:complete} and][]{Frayer2004}.  We are reassured by the range in values we find for the ratio between these SFRs are consistent with those found in \citet{Daddi07}, who studied BzK-selected galaxies between $1.4<z<2.5$ in the GOODS fields, and who also compared infrared and UV star formation rate indicators.    

Since the main conclusions we draw in this work are based on the mass-complete GNS sample using UV-based SFRs, and the corrections derived to apply to the 24 $\mu$m data are still new, we do not attempt to make any correction to the IR-based SFRs in this study.

\section{Results}\label{sec:results}

In this section, we discuss the star-forming properties of GNS galaxies as a function of stellar mass and rest-frame $(U-B)$ colour.  We discuss the effects of the dust determination and the photometric redshift errors on these relations.   We also investigate what properties may account for the red colours observed in galaxies at high redshift.

\subsection{Star Formation and Stellar Mass}\label{sec:sfr_mass}

\begin{figure*}
\includegraphics[width = 0.9\textwidth]{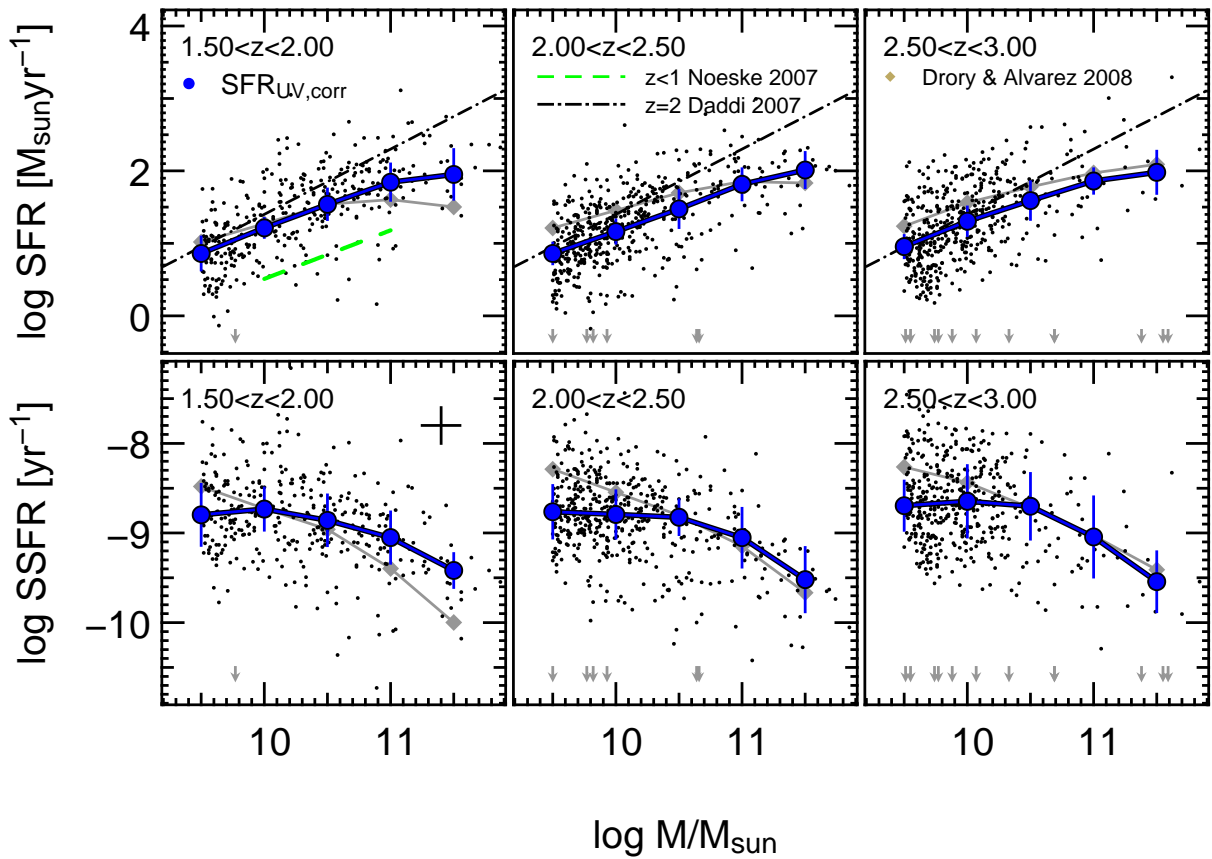}
\caption{\label{sfr_ssfr_mass}
Median SFR$_{UV,corr}$ versus stellar mass in three redshift bins (top) and median SSFR versus stellar mass (bottom).  Median SFRs are shown for GNS galaxies in each stellar mass bin as large blue circles.  Individual GNS galaxies are black points, with galaxies below the $z_{850}$-band detection limit shown as grey arrows.   Grey diamonds show results from \citet{drory_alvarez} who use SED fitting to derive UV SFRs.  The dash-dot diagonal line in the top row is the \citet{Daddi07} relation for galaxies at $z\sim2$ while the green dashed line shows the intermediate relation found by \citet{noeske07} at $z\sim0.8$.
}
\end{figure*}

Several studies in recent years have investigated the change of SFR as a function of stellar mass and time to redshifts of $z\sim1.5$ \citep[e.g.][]{BE00,Bauer05,Bundy06,Zheng07,noeske07} and have found a general trend for SFRs to rise with increasing stellar mass.  Studies of galaxies at higher redshift, between $1.5<z<2.5$, using SFR indicators from many different wavelength regions are also beginning to show evidence for increasing SFRs with stellar mass \cite[e.g.][]{Daddi07,Renzini2009,Santini2009,Onodera2010}.  We present here for the first time the SFR-stellar mass relation for a sample complete in stellar mass to $M_{*}=3\times10^{9} \Msun$ between $1.5<z<3$.

Figure~\ref{sfr_ssfr_mass} shows SFR$_{UV,corr}$ as a function of stellar mass in three redshift bins, from $z=1.5$ to $z=3$.  It is clear that a trend between SFR and stellar mass exists at each of these redshifts.    Within a single redshift bin, the median SFR increases by roughly a factor of 10 from the lowest stellar mass galaxies probed, log~$M_{*}= 9.5$, to the highest mass, log~$M_{*}\sim12$ galaxies.  Galaxies not detected at rest-frame UV wavelengths are considered non-star-forming.   Non-SF galaxies compose only 1.5\% of the sample and are shown as downward facing grey arrows in Figure~\ref{sfr_ssfr_mass}.  In computing the median SFR$_{UV,corr}$ values (large blue circles), we include galaxies not detected at rest-frame UV wavelengths in each mass and redshift bin.  

The correlation between stellar mass and SFR for the GNS sample has a slope consistent with that found at intermediate redshift by \citet{noeske07}, shown as the green dashed line in Figure~\ref{sfr_ssfr_mass}.  Using the same stellar mass range as \citet{noeske07} ($10<~$log$~M_{*}<11$), the normalisation found for GNS galaxies at $z\sim2$ is a factor of 2.2 larger than at $z\sim0.7$.  At fixed stellar mass, star-forming galaxies are more actively forming stars, on average, at $z\sim1.5$ than at $z\sim0.7$, likely due to a larger abundance of gas that depletes with time \citep{noeske07}.  

In a study of $K$-selected GOODS galaxies at $z\sim2$, \citet{Daddi07} also find increasing SFRs with stellar mass, as shown by the dash-dot line in the top row of Figure~\ref{sfr_ssfr_mass}.  The correlation of the GNS sample has a flatter slope than the \citet{Daddi07} relation at $z\sim2$, but similar median SFR values for log~$M_{*}<11$ galaxies.  Among massive log~$M_{*}\ge11$ GNS galaxies, we find an average SFR$_{UV,corr} = 103\pm75~\Msun~$yr$^{-1}$, which is roughly a factor of 1.6 less than the \citet{Daddi07} relation.  We find a flattened relation relative to \citet{Daddi07} which is either due to the fact that we are using a mass-complete sample instead of just star-forming galaxies, or due to the overestimation of the dust correction applied by \citet{Daddi07}, as found by \citet{Nordon2010} and discussed in Section~\ref{sec:dust} above.    Interestingly, for a given stellar mass, we see roughly constant SFRs for all galaxies over the redshift range $1.5<z<3$ for the full mass-selected GNS sample.

\begin{figure*}
\includegraphics[width = 0.8\textwidth]{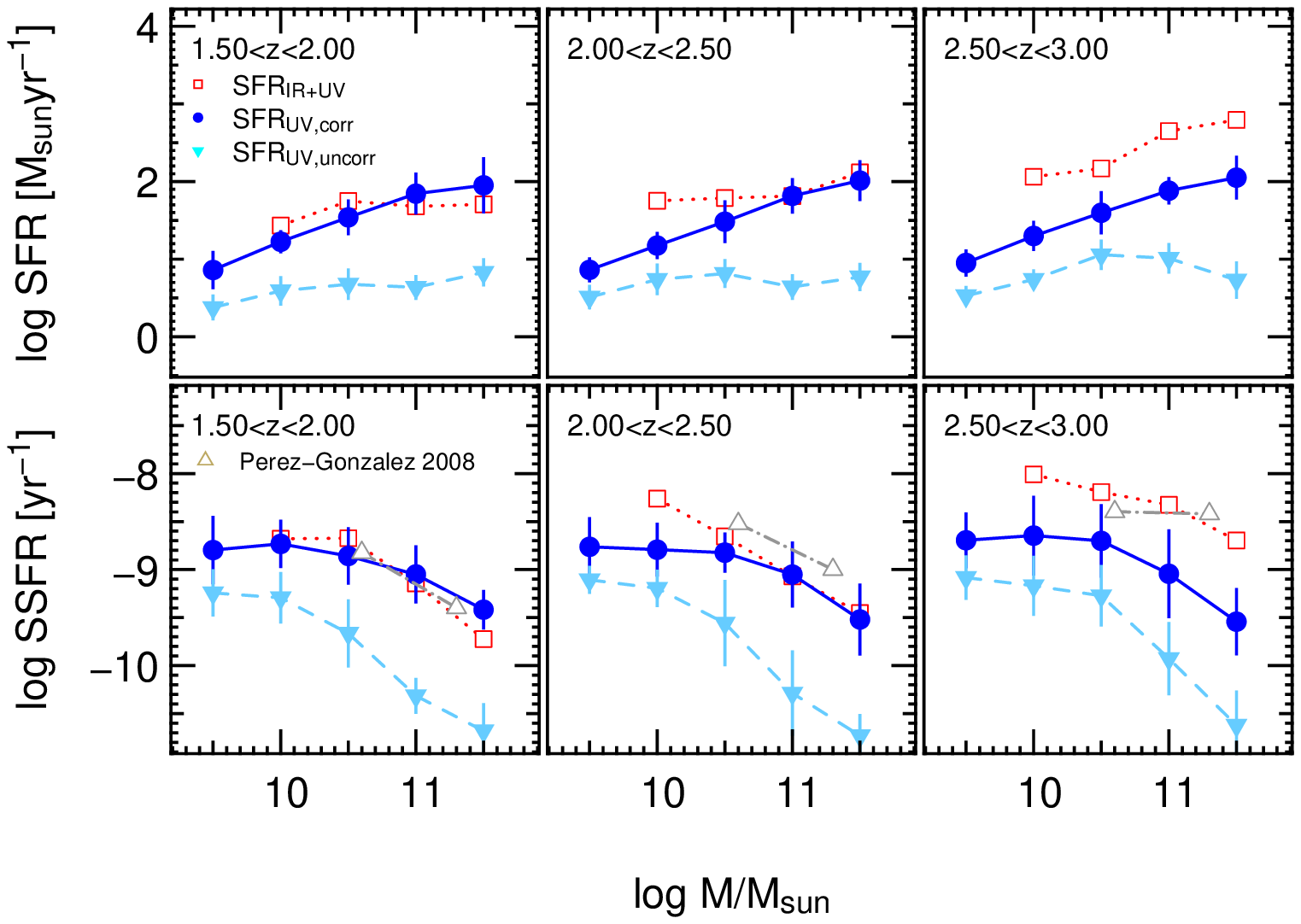}
\caption{\label{sfr_ssfr_mass_avg}
Median SFR versus stellar mass in three redshift bins (top) and median SSFR versus stellar mass (bottom).  The blue points are the same SFR$_{UV,corr}$ as in Figure~\ref{sfr_ssfr_mass}.  Light blue inverted triangles show SFR$_{UV,uncorr}$ for all GNS galaxies.   Open red squares are median SFR$_{IR+UV}$ for GNS galaxies detected at 24$\mu$m.   Open gray triangles show SFR$_{IR}$ from Perez-Gonzalez et al. (2008).
}
\end{figure*}

We plot the 19 non-star-forming galaxies as grey arrows in Figure~\ref{sfr_ssfr_mass}.  Recall from Section~\ref{sec:sfr} that non-star-forming galaxies are those not detected at the observed ACS/$z_{850}$-band 5~$\sigma$ limit of 27.5 in the AB system, corresponding to an uncorrected SFR$_{UV}$ of $0.98\pm0.3~ \Msun$yr$^{-1}$ at $z=3$.  Only one non-star-forming galaxy is at $z<2$ and only four are detected at 24 $\mu$m, implying that we are not missing a significant population of highly extincted starbursts at $z>2$.  The non star-forming galaxies are mostly low mass, and only three have $M_{*}\ge10^{11}\Msun$.  We will look more at the colours of these galaxies later, in Section~\ref{sec:sfr_colour}.

We present the specific star formation rate (SSFR = SFR/M$_*$) as a function of stellar mass in the bottom three panels of Figure~\ref{sfr_ssfr_mass}.  The error bars show the standard deviation of the distribution of galaxies in each bin.  The median SSFR of galaxies with log~$M_{*}~\le~10.5$ is largely independent of stellar mass and redshift over the $\sim2$~Gyr period between $z=3$ and $z=1.5$.   Galaxies with log~$M_{*}~ \ge~10.5$ show a sharp decrease in SSFR with increasing stellar mass such that the median SSFRs decrease by 0.6 to 1.0~dex between log~$M_{*}~\sim~10.5$ and log~$M_{*}~\sim~11.5$.  

Using radio observations of SSFRs, \citet{Pannella2009_sfr} also find that the average SSFR is constant over almost one dex in stellar mass at $z=1.6$ and $z=2.1$.  \citet{Dunne09} also find a shallow decrease of SSFRs with decreasing stellar mass using stacked radio observations.  As a function of redshift, we see no dependence of SSFR with redshift, in contrast to studies at lower redshift \citep{BE00,Bauer05,PerezGonzalez05,Zheng07,Pannella2009_sfr}, but in agreement with recent results from the infrared using the Hershel Space Telescope \citep{Rodighiero2010,Cava2010}.  

It is possible that the low SSFRs we see at $z\sim3$ for high mass galaxies is indicative of a period of stellar mass growth dominated by merging events, which subsequently slow at $z<3$ so that star formation becomes the dominant growth mechanism for the most massive galaxies until their cold gas supply runs out.   A massive galaxy shown in Figure~\ref{sfr_ssfr_mass} with a SSFR between the median values of logSSFR~$=-9.0$~yr$^{-1}$ and $-9.5$~yr$^{-1}$ will double its stellar mass from star formation alone in 1 to 3 Gyr.

\begin{figure*}
\includegraphics[width = 0.8\textwidth]{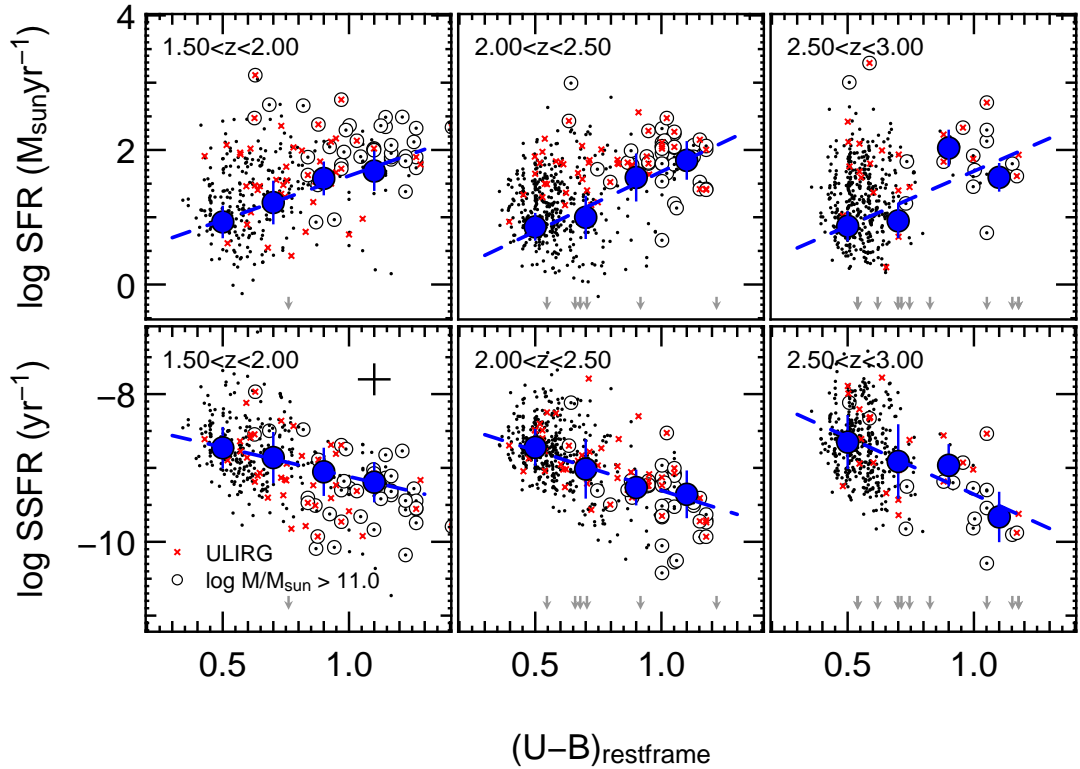}
\caption{\label{sfr_colour}
Star formation rate versus rest-frame $(U-B)$ colour in three redshift bins (top) and specific star formation rate versus $(U-B)$ colour (bottom).  Red crosses show galaxies classified as ULIRGs (L$_{IR}  \ge 10^{12}$\Lsun) and open circles identify massive galaxies with $M_{*} > 10^{11} \Msun$.  }
\end{figure*}

\subsection{The Effect of Photometric Redshifts}\label{mc}

The errors on the photometric redshifts determined for GNS galaxies are consistent with those found in the literature for galaxies in this redshift range for studies using a similar number of filter bands, as described in Section~\ref{sec:redshifts}, but in order to further test the effect of the photometric redshift errors on our results, we have performed a set of Monte Carlo simulations (fully described in \citet{Grutzbauch2011}).  We randomise the photometric redshift input according to the typical photometric redshift error in our redshift range of $\Delta z/(1+z) \sim 0.1$, obtained by the comparison to the available secure spectroscopic redshifts for the galaxies in our sample.  

We assume a Gaussian distribution of errors on the photo-z redshift for each object, where the width of the distribution corresponds to $\Delta z/(1+z) \sim 0.1$ for all galaxies in the redshift range $1.5<z<3$.  For each galaxy, a random value is selected within this distribution, which is then added to the measured photo-z. Galaxies outside our redshift range of interest are included in the randomisation to account for scattering in and out of our redshift range.  Catastrophic outliers are accounted for by randomly adding much larger offsets to the original redshifts, of the percentage of galaxies corresponding to the fraction of catastrophic outliers, which is calculated from the comparison with secure spectroscopic redshifts.  

We then repeat the computation of the SFR and SSFR as a function of stellar mass and colour with the new photometric input.  The results indicate that while the individual values of SFR and SSFR for a single galaxy can change significantly, on average, the relations we report in the Section~\ref{sec:sfr_mass} and later in Section~\ref{sec:sfr_colour} remain robust and unchanged despite the errors of the photometric redshifts used.  

In order to determine how much the lower photometric redshift reliabilities of fainter galaxies affect the results presented here, we repeat the Monte Carlo simulations using a magnitude-dependent approach.  We calculate the photometric redshift error and outlier fraction separately for the bright and faint galaxies separating them at $H = 24$ mag. The main impact of this procedure is an increase in the catastrophic outlier fraction for the faint galaxies in our redshift range to 33\%, as opposed to 17\% for the brighter galaxies. The random error stays about constant for both samples at $\Delta z/(1+z) \sim 0.1$, whereas the offset increases slightly from $\sim$0.05 for the brighter sample to $\sim$0.08 for the fainter galaxies.  

The result of repeating the simulations with the above values separately for bright and faint galaxies is that the number of galaxies in each bin decreases due to the large outlier fraction which scatters one third of the faint galaxies out of the redshift range of this study entirely, but we find that the overall results remain unchanged. We do not correct the measured values as a result of the Monte Carlo simulations, because we have found that the uncertainty around the resulting relations is hardly affected, other than the scatter increases slightly.  While the uncertainty is larger for less massive galaxies, the results are robust to various photometric redshift errors and improved reliability will be a goal of future spectroscopic redshifts, but that these will have to be done with 20-30m telescopes due to the faintness of these systems.

\subsection{The Effect of Dust}

In order to see how the influence of dust affects these results, we show several different measures of SFR in Figure~\ref{sfr_ssfr_mass_avg}, which shows the SFR and SSFR as a function of stellar mass.   The blue circles are the same SFR$_{UV,corr}$ presented in Figure~\ref{sfr_ssfr_mass}.  Light blue inverted triangles show the SFR$_{UV,uncorr}$ before a dust correction is applied.  The SFR$_{UV,uncorr}$ are consistent with a slight increase with stellar mass in each redshift bin, and also a slight increase in SFR as a function of redshift in each stellar mass bin, except for the most massive galaxies which have the same median values at all redshifts.  This implies that the dust correction is a function of stellar mass \citep{Grutzbauch2011} and not only steepens the SFR-M$_{*}$ relation but also flattens the SSFR-M${*}$ relation.  

The open red squares in Figure~\ref{sfr_ssfr_mass_avg} show the median SFR$_{IR+UV}$ for GNS galaxies detected at 24~$\mu$m.  At $z<2.5$, the SFR$_{IR+UV}$ and SFR$_{UV,corr}$ agree quite well.  On the other hand, at $z>2.5$ the SFR$_{IR+UV}$ is greater than SFR$_{UV,corr}$ by an average factor of 5 and greater than SFR$_{UV,uncorr}$ by an average factor of 50.  A high redshift infrared study by \citet{pg_2008_z4} uses {\it Spitzer Space Telescope} observations at 24$\mu$m to determine SFRs of galaxies at $z\sim2$.  Their results for log~$M_{*}~\ge~10.5$ galaxies are shown in the bottom panels of Figure~\ref{sfr_ssfr_mass_avg} as grey triangles connected by a dash-dot line.   At $z<2$, the SSFR from \citet{pg_2008_z4} are in excellent agreement with both SSFR$_{IR+UV}$ and SSFR$_{UV,corr}$ of the GNS sample.   As redshift increases, \citet{pg_2008_z4} find higher values of mid-IR-derived SSFR and by $z>2.5$, both measurements of mid-IR-derived SSFR are one dex higher than the SSFR$_{UV,corr}$ value and nearly two dex higher than SSFR$_{UV,uncorr}$.  

Based on the overall agreement between the median dust-corrected SFR$_{UV}$ and the mid-infrared derived SFR at $z<2.5$ shown in Figure~\ref{sfr_ssfr_mass_avg}, it appears we are successfully able to recover the dust obscured star formation using the rest-frame UV slope method to determine the dust extinction.   The large difference at $z>2.5$, however, is likely a combination of the SFR$_{IR+UV}$ being exceptionally high at $z>2.5$  \citep[e.g.][]{Papovich2007,Elbaz2010}, the SFR$_{UV,corr}$ representing the full mass-selected sample, including galaxies not detected at 24~$\mu$m, and the uncertainty in using the UV spectral slope to determine UV dust extinction at these high redshifts.

\subsection{Star Formation and Restframe Colour}\label{sec:sfr_colour}

\begin{figure*}
\includegraphics[width = 0.95\textwidth]{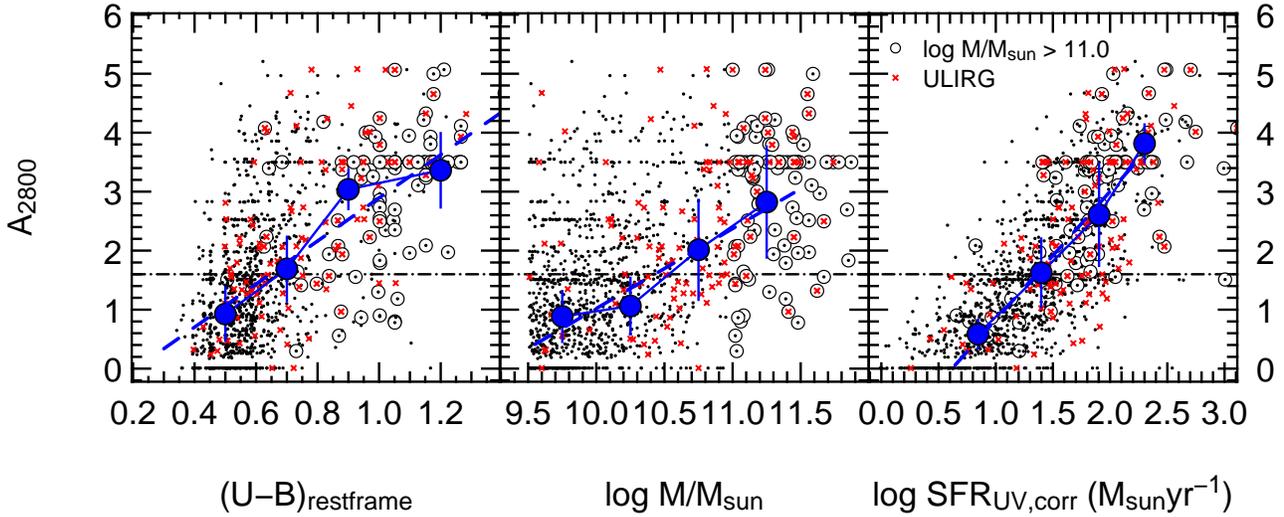}
\caption{\label{dust}
Reddening factor A$_{2800}$ as a function of rest-frame $(U-B)$ colour (left), stellar mass (center), and SFR$_{UV, corr}$ (right).   The average A$_{2800}$ value for all GNS galaxies is shown by the dash-dot line.  The open circles show galaxies with log~$M_{*}~ \ge~11$ while the red crosses identify galaxies classified as ULIRGs.}
\end{figure*}

Based on optical and near-infrared colours of galaxies, there is evidence that massive galaxies with red colours at $z\sim2$ are probable progenitors of red non-star-forming massive galaxies found at low redshift \citep[e.g.][]{Conselice2007,Kriek2008}.  Usually, it is assumed that red massive galaxies seen in the distant universe consist of old stellar populations which passively evolve until the present day.  However, our results show that it is possible that these massive galaxies are red due to dust, or a combination of older stellar populations and dusty star formation, as we investigate in this section.

One way to identify galaxies with large amounts of dusty star formation is to look at populations with excess infrared emission such as ULIRGs, classified as galaxies with log ($L_{IR}$/~\Lsun) $> 12$.   The infrared continuum of a ULIRG is dominated by thermal re-radiation by dust.  Dust particles absorb UV light, mostly emitted by massive main-sequence stars in regions of intense star formation activity, and re-radiate the energy at longer wavelengths.  We saw in Figure~\ref{irx_beta} that ULIRGs have excess IR luminosity relative to UV luminosity, and we investigate here whether these galaxies have higher dust-corrected SFRs, and from what their rest-frame $(U-B)$ colours might originate. 

In Figure~\ref{sfr_colour}, we present SFR$_{UV,corr}$ (top) and SSFRs (bottom) as a function of rest-frame $(U-B)$ colour, split into three redshift bins.  Symbols are the same as in Figure~\ref{sfr_ssfr_mass} with the addition of red crosses indicating galaxies detected at 24$\mu$m that are classified as ULIRGs.  It is clear from Figure~\ref{sfr_colour} that low-mass galaxies have bluer colours with the majority exhibiting rest-frame $(U-B)~<~$0.8.  Massive galaxies  are almost all red with rest-frame $(U-B)~>~$0.8. 

We find that galaxies classified as ULIRGs span the entire range of rest-frame $(U-B)$ colours but show a trend for exhibiting higher SFRs and SSFRs at $z>2$.  Galaxies classified as ULIRGs are believed to be the most common mode of star formation in massive galaxies \citep[e.g.][]{Daddi07}.  The blue dashed line in Figure~\ref{sfr_colour} shows the 2 $\sigma$ clipped least squares fit to all the galaxies in the mass-complete GNS sample.  In the upper panels showing the SFR, we find that 80\% of the 50 ULIRGs in the middle redshift bin of $2<z<2.5$ lie above the least squares fit.  The fraction is 59\% in the lowest redshift bin and 70\% in the highest.  We see a general trend in the lower panels of Figure~\ref{sfr_colour} where SSFRs decrease with redder colours, as seen previously at $z<1.5$ \citep{Cooper2008,Salim09}.  All the galaxies with the highest SSFRs have blue rest-frame $(U-B)$ colours and lower stellar masses.  This is evidence for the presence of dusty star formation in galaxies of all stellar masses above the 24~$\mu$m detection limit imposed here.

\subsection{What Causes $z>1.5$ Galaxies to be Red?}\label{sec:redness}

We found in previous sections that massive galaxies are redder, and form stars at higher rates than lower mass galaxies.  In order to understand what dominates the cause of the red colours of these galaxies, we look more in depth at the red galaxy population.  We define red galaxies to be those with rest-frame $(U-B)\ge0.8$ which roughly divides the sample into a blue cloud and a red sequence (see Figure~\ref{sfr_colour} and Grutzbauch et al. 2011).  From the complete mass-selected GNS sample, we find that $\sim14\%~\pm$~1\%~(184/1282) are red according to this criterion, and among these red galaxies, the vast majority, $\sim88\%~\pm$~7\%~(163/184) have log~$M_{*}~ \ge~10.5$.  Since massive galaxies dominate the red population, and because the 24$\mu$m detection fraction drops drastically to less than $\sim$25\% for galaxies with log~$M_{*}~<~10.5$ (see Section~\ref{sec:lir} and Figure~\ref{f24}), we focus the results in this section on log~$M_{*}~\ge~10.5$ galaxies. 

In order to distinguish {\em dusty} red galaxies, as opposed to galaxies that are red due to evolved stellar populations, we use three distinguishing characteristics of the galaxy population:  (1) the level of dust extinction based on A$_{2800}$, (2) the infrared properties, and (3) star formation rates. 

First, we examine the dust absorption of UV light within red galaxies based on the A$_{2800}$ values derived from the SED-determined rest-frame ultraviolet slope.  Using this method, we denote {\em dusty} red galaxies as those that have extinction values greater than the average for all galaxies.  The average extinction value, A$_{2800}=1.6$, is shown as the horizontal dash-dot line in Figure~$\ref{dust}$.   In this way we find that among log~$M_{*}~ \ge~10.5$ red galaxies, a high fraction, $\sim87\%~\pm$~7\%~(142/163), are red due to dust, and a similarly high fraction for log~$M_{*}~\ge~11$ galaxies: 89~$\pm$~7\%.  

Galaxies with red colours and A$_{2800} \le 1.6$ are not considered dusty based on this prescription and therefore likely derive their red colour from old stellar populations.  This amounts to $\sim13\%~\pm$~3\%~(21/163) of the log~$M_{*}~ \ge~10.5$ galaxies and $\sim11\%~\pm$~4\%~(10/90) among the most massive log~$M_{*} \ge11$ galaxies.  In this population of massive red, non-dusty galaxies, half (5) are curiously detected at 24 $\mu$m and two of those are classified as ULIRGs, possibly due to dust heating by another source such as an AGN \citep{Bluck2011}.     

Dust attenuation increases as a function of stellar mass as shown in the middle panel of Figure~$\ref{dust}$.  Using the COSMOS survey, \citet{Pannella2009} publish a relation between A$_{1500}$ and stellar mass.  Assuming the Calzetti dust law to convert their relation to A$_{2800}$, we find that the extinction values agree at log~$M_{*}~=~10.2$, but their relation is a steeper function of mass such that their A$_{2800}$ at  log~$M_{*}~\ge~11$ is higher than the one found in this work by roughly one magnitude.  This difference is true for the full GNS mass-selected sample or just galaxies restricted to the redshift range of $1.5<z<2$.  This could be due to the fact that in this study we assign extinction values of A$_{2800} = 3.5$ to galaxies with $\beta > 1.0$ (60 objects)in order to avoid overcorrecting for dust (see Section~\ref{sec:dust}).  

Next, we identify dusty red galaxies as those detected at 24$\mu$m and classified as ULIRGs, since detection at 24$\mu$m implies that a galaxy has a significant amount of dust.  In terms of 24$\mu$m detections alone, the maximum percentage of non-dust-dominated red galaxies with log~$M_{*} \ge10.5$ corresponds to the number of red massive galaxies (163) not detected at 24$\mu$m, which is $45\%~\pm$~5\%~(74/163).  This percentage is quite different than the value of $\sim$13\% found using the extinction criterion described above.  Placing significance on whether a galaxy is classified as a ULIRG is slightly arbitrary, but at $z\sim2$, ULIRGs are proposed to be a common mode of star formation in massive galaxies \citep[e.g.][]{Daddi07}.  We find that among the 24~$\mu$m detections of massive red galaxies, $52\%~\pm~6\%$~(47/89) are classified as ULIRGS indicating that they harbour significant amounts of dust, and reassuringly, 92\% of those also have above average values of A$_{2800} = 1.6$.  

As shown in Figure~\ref{f24}, higher fractions of galaxies are detected at 24 $\mu$m in the redshift range $1.5<z<2$ since we are possibly only detecting the most extreme infrared objects at $z>2$.   If we do the same exercise examining dust in galaxies based on infrared properties with just galaxies between $1.5<z<2$, we find that the maximum percentage of non-dust-dominated red galaxies with log~$M_{*} \ge10.5$ corresponds to the number of red massive galaxies (79) not detected at 24$\mu$m is $41\%~\pm$~4\%~(32/79), very similar to the value found for the full redshift range.  The fraction of ULIRGS among the low redshift massive red galaxy population is lower than for the full redshift range at $43\%~\pm~3\%$~(20/47).

Despite the red colours of massive galaxies, the average SFR$_{UV,corr}$ of red galaxies with log~$M_{*}~\ge~10.5$ is $78~\pm~58~\Msun~$yr$^{-1}$, consistent over the entire redshift range we probe.  Recall that among the full GNS sample, we find that 1.5\% are not detected in the $z_{850}$-band and are therefore considered non-star-forming.   Among massive red galaxies, we find that $3.6\%~\pm~1.4\%$~(6/163) are not detected in the observed $z_{850}$-band.  If we look to the bluer wavelength of the observed $B$-band (rest frame UV), we find that 87.5\% of the total sample is not detected at the 5$\sigma$ level.  Among massive red galaxies,  $42\%~\pm~5.0\%$~(68/163) are not detected in $B$-band and yet massive galaxies have the highest SFRs in the sample and still have red colours, as shown in Figure~\ref{sfr_colour}.  If we assume that dust absorption is what causes the lack of detection in the $B$-band \citep[][see also Section~\ref{sec:dust}]{Daddi2004}, then we can conclude that between 45\% and 85\% of massive red galaxies harbour evidence of dusty star formation.  If we were overestimating the derived value of dust extinction for any of the galaxies in the sample, the effect would be to slightly lower this percentage of massive red galaxies with dusty star formation.

We can compare these results to previous work at lower redshifts.  Previously, it has been shown that extinction contributes to the red colours of galaxies at $z=1$.   For example, \citet{Smail2002} study a complete sample of 68 EROs with $(R-K)\ge5.3$ and $K<20.5$.  Using radio maps and photometric classifications, they find that between 30\% and 60\% of the ERO population are dusty, star-forming galaxies, as opposed to passively evolved stellar systems.  The K-selected K20 Survey finds among the ERO population at $z\sim1$ that the proportion of dusty starbursts at $K<19.2$ is between 33\% - 67\% \citep{K20-02}.   Recent results from the large area COSMOS survey also indicate that half of the nearly 5000 EROs in their sample are dusty starburst galaxies \citep{Kong2008} based on three different methods of classification: SED fitting, infrared colour, and morphology.  Furthermore, \citet{Whitaker2010} study massive (log~$M_{*}~ \ge~11$) galaxies in the NEWFIRM Medium-Band Survey to $z\sim2.2$ and find a large scatter in the colours of quiescent galaxies.  This scatter in rest-frame $(U-V)$ colours of quiescent galaxies arises from the spread in ages, due to the presence of both relatively quiescent red, old galaxies and quiescent blue, younger galaxies toward higher redshift.

The conclusion is that while massive galaxies have on average higher SFRs than lower mass galaxies, they also have dustier star formation regions at $z>1.5$, causing their overall colours to be redder than lower mass galaxies.  One possibility is that we may be seeing evidence for massive galaxies producing more high mass stars due to having a more top-heavy initial mass function (IMF).  It has been suggested that discrepancies between the global star-formation history and the stellar mass history can be reconciled by using evolving IMFs, such that more top-heavy IMFs are more common within star forming galaxies at $z>2$ \citep[e.g.][]{pg_2008_z4,Wilkins2008,dave2008}.  The most massive stars, i.e., Wolf-Rayet stars create dust, and therefore more dust would be produced in top-heavy IMF star-forming regions.   Another possibility is simply that massive galaxies are able to retain more of the dust from high mass star formation due to their large gravitational potentials, regardless of the IMF.  Although, it is possible that if we used a more top-heavy IMF to determine stellar masses, some fraction of these galaxies would no longer be considered massive ($M_{*} \ge10^{11}\Msun$) in our sample.

\section{Conclusions}\label{sec:conclusions}

We present the first results of a study designed to investigate the star-forming properties of a stellar mass selected sample of galaxies in the GOODS NICMOS Survey (GNS) with masses log~$M_{*} \ge9.5$ in the redshift range $1.5<z<3.0$.  The GNS is based on deep Hubble Space Telescope F160W (NICMOS/H-band) imaging of the GOODS North and South fields.  We combine the GNS NICMOS imaging with HST/ACS and {\em Spitzer} GOODS photometry to calculate photometric redshifts, stellar masses, rest-frame colours, dust extinctions, and star formation rates from the ultraviolet and infrared.  Our sample includes $\sim$1300 galaxies with log~$M_{*} \ge9.5$ between $1.5<z<3$, of which eight percent are massive galaxies with $M_{*} \ge10^{11}\Msun$.  

We measure both UV and IR based star formation rates for these galaxies.  We determine SFR$_{UV}$ from the observed optical ACS $z_{850}$-band flux density which corresponds to the rest-frame UV luminosity around 2800$\mathrm{\AA}$.  We find that the best way to determine the amount of dust extinction uniquely for each galaxy is to calculate the ultraviolet slope from UV luminosities calculated from multi-wavelength SED-fitting.  The average value of A$_{2800}$ across the entire sample is $1.6~\pm~1.2$~mag.  We find that 21\% of the sample are detected at 24$\mu$m above the 30 $\mu$Jy (5$\sigma$) limit, and that the detection fraction is much higher, $\sim60\%$, for log~$M_{*} \ge 11$ galaxies.  

The dust-corrected UV star formation rates (SFR$_{UV,corr}$) for all GNS galaxies cover a range between 0.6 $\Msun$yr$^{-1}$ and $\sim$2000 $\Msun$yr$^{-1}$.  The average SFR$_{UV,corr}$ for the full mass-selected sample is $42\pm53~\Msun$yr$^{-1}$, and for just the massive galaxies ($M_{*}\ge10^{11}\Msun$), we calculate an average SFR$_{UV,corr}$ of $103\pm75~\Msun$yr$^{-1}$, consistent across the entire redshift range.  We compare SFR$_{UV,corr}$ with SFR$_{IR+UV}$ determined from the IR luminosity at 24$\mu$m added to the uncorrected UV derived star formation rates.  Among massive galaxies, $\sim$87\% agree in SFR using these two methods to within a factor of 10, while 80\% of all the galaxies in Figure~\ref{sfr_uvIR} agree to within a factor of 10.  On average, SFR$_{IR+UV}$ is a factor of 3.5 larger than SFR$_{UV,corr}$ between $1.5<z<3$, with a scatter of 4.6. 

We find that SFRs increase with stellar mass over the full redshift range, but the average SFR at a given stellar mass remains constant over this $\sim2$ Gyr period, opposite to what is seen at $0<z<1$ when massive galaxies have a declining SFR.  This implies that whatever effect produces the downsizing seen in massive galaxies at $z < 1$ has not yet, or has not yet had enough time, to significantly decrease the star formation rate within these systems at $1.5 < z < 3.0$.  

Furthermore massive galaxies (log $M_{*} \ge11$) in our sample have the highest SFRs, yet exhibit red rest-frame $(U-B)$ colours at all redshifts between $1.5<z<3$.  We conclude that between $\sim$45-85\% of massive galaxies at these redshifts are undergoing dusty star formation, despite a higher fraction ($>$85\%) exhibiting red colours.  We suggest that the red colours of massive galaxies is evidence for these systems having dustier star formation regions than galaxies with lower stellar mass.  One cause for this could be that high mass galaxies may have more top-heavy IMFs than lower mass galaxies, thus producing more Wolf Rayet stars which are responsible for producing dust.  

\section*{Acknowledgments}

We acknowledge support from the UK Science and Technology Facilities Council (STFC), the Leverhulme Trust, and the University of Nottingham.  PGP-G acknowledges support from the Spanish Programa Nacional de Astronom\'{\i}a y Astrof\'{\i}sica under grants AYA2009Ð07723ÐE, AYA 2009--10368 and CSD2006-00070, and the Ram\'on y Cajal Program financed by the Spanish Government and the European Union.

\bibliographystyle{mn2e}
\bibliography{bauer_gns}

\bsp

\label{lastpage}

\end{document}